\documentclass[a4paper]{article}

\title{Study of movement coordination in human ensembles via a novel computer-based set-up}
\date{}

\author{
Francesco Alderisio\footnotemark[2] \footnotemark[4] \ , Maria Lombardi\footnotemark[3] \footnotemark[4] \ , Gianfranco Fiore\footnotemark[2] \ \& Mario di Bernardo\footnotemark[2] \footnotemark[3] \ \text{*}
}

\usepackage{amsmath}
\usepackage{amssymb}
\usepackage{amsthm}
\usepackage{hyperref}
\usepackage{graphicx}
\usepackage{subfig}
\usepackage{epstopdf}
\usepackage{booktabs}
\usepackage{tabularx}  
\usepackage{xfrac}
\usepackage{color}
\usepackage{lineno}
\usepackage[margin=0.8in]{geometry}

\begin{document}

\begin{titlepage}

\maketitle
Movement coordination in human ensembles has been studied little in the current literature. In the few existing experimental works, mostly situations where all subjects are connected with each other through direct visual and auditory coupling, and unavoidable social interaction affects their coordination level, have been investigated. Here, we study coordination in human ensembles via a novel computer-based set-up that enables individuals to coordinate each other's motion from a distance so as to minimize the influence of any other form of social interaction. The proposed experimental platform makes it possible to implement different visual interaction patterns among the players, so that participants can only take into consideration the motion of a designated subset of the others. This allows the evaluation of the exclusive effects on coordination of the structure of interconnections among the players in the group and their own dynamics.
In order to further analyze the mechanisms underlying human coordination, our set-up enables the deployment of virtual computer players to investigate dyadic interaction between a human and a virtual agent, as well as group synchronization in mixed teams of human and virtual agents.
We use this novel set-up to study coordination both in dyads and in groups over different structures of interconnections, with and without virtual agents acting either as followers or as leaders. We find that, in dual interaction, virtual players manage to interact with participants in a human-like fashion, thus confirming findings in previous work showing that virtual agents succeed in reproducing the kinematic features characterizing human motion. We also observe that, in group interaction, the level of coordination among humans in the absence of direct visual and auditory coupling depends on the structure of interconnections among participants. This confirms, as recently suggested in the literature, that different coordination levels are achieved over diverse visual pairings among participants in the presence as well as in the absence of social interaction.
Finally, we present preliminary experimental results on the effect on group coordination of deploying virtual computer agents in the human ensemble.

\footnotetext[2]{Department of Engineering Mathematics, Merchant Venturers Building, University of Bristol, Woodland Road, Clifton, Bristol BS8 1UB, United Kingdom (\texttt{enmdb@bristol.ac.uk})}
\footnotetext[3]{Department of Electrical Engineering and Information Technology, University of Naples Federico II, Via Claudio 21, 80125 Naples, Italy (\texttt{mario.dibernardo@unina.it})}
\footnotetext[4]{These authors have contributed equally as first authors}

\end{titlepage}

\setcounter{page}{2}

\section{Introduction}
Interpersonal coordination between the motion of two individuals performing a joint task has been extensively studied over the past few decades \cite{schmidt1994phase,richardson2007rocking,oullier2008social,schmidt2008dynamics,marsh2009social,varlet2011social,walton2015improvisation,slowinski2016dynamic}; a recent example being that of the \emph{mirror game}, presented as paradigmatic case of study where human participants (HP) imitate each other'€™s movements in a pair \cite{noy2011mirror}.
In general, multiplayer scenarios have been investigated less than those involving only two participants, because of practical problems in running the experiments and the lack of models accounting for movement coordination in human groups, in contrast to the numerous studies dealing with animal groups \cite{couzin2005effective,nagy2010hierarchical,nagy2013context,zienkiewicz2015leadership}.

Some of the existing results on multiplayer human coordination include studies on rocking chairs \cite{frank2010test, richardson2012measuring,alderisio2016entrainment}, group synchronization of arm movements and respiratory rhythms \cite{codrons2014spontaneous}, music \cite{glowinski2013movements,badino2014sensorimotor,volpe2016measuring} and sport activities \cite{wing1995coordination,yokoyama2011three}.
In these papers, the behavior of a group of people performing some coordinated activities is analyzed, but the features and the level of coordination are not explicitly correlated to the way the players interact (i.e., the structure of their connections).
In general, in the experiments reported in the literature all subjects involved share direct visual and auditory coupling with each other; moreover, inevitable social interaction affects the level of coordination in the group \cite{healey2005inter,kauffeld2009complaint,passos2011networks,d2012leadership,duarte2012intra,duarte2013influence,glowinski2013movements,cardillo2014evolutionary}.
Indeed, body movements, friendship relationships, shared feelings, particular affinities and levels of hierarchy might have a significant impact on how each individual in the ensemble chooses her/his preferred partner(s) to interact the most with \cite{baumeister1995need,mas2010individualization,stark2013generalization}.
This makes it difficult to assess the impact on the group coordination level of solely varying the structure of interconnections among its members, a phenomenon that has been suggested as crucial in determining the level of coordination arising in a human group \cite{passos2011networks,duarte2012intra,duarte2013influence,alderisio2016interaction}.

To address this  problem and study the emergence of human coordination in the absence of unintentional social interaction, we present here ``Chronos'', a novel computer-based set-up that allows participants to perform a joint task from a distance, both in dyads and in groups.
In our platform, each subject runs a serious computer game where s/he can move the position of an object on her/his computer screen and see traces of the objects moved by the other players. The software makes it possible to show on each screen only the traces of a designated subset of the players in the group as decided by an Administrator, so that different visual interaction patterns among its members can be implemented.
To prevent any other form of interaction, participants are visually separated when in the same environment by barriers, and hear white noise (through headphones connected to their computers) isolating them from any external sound. 
(Alternatively, players can join the group remotely.)
Therefore, subjects have no information on the identity of those they are interacting with and receive no direct visual or behavioral cues from other members in the group.

In order to further analyze the mechanisms underlying human coordination and explore features that are not easily accessible in ordinary human interactions, we take inspiration from the \emph{human dynamic clamp} introduced in \cite{dumas2014human} and extend it to a multiplayer scenario. Indeed, our computer-based architecture allows the trace of some objects on the players' screen to be moved by virtual players (VP) driven by a computational cognitive architecture as described in the rest of the paper. 

To illustrate and validate the set-up, we use it to investigate how the structure of the interconnections among players in a group of $5$ persons affects their coordination level. We show that the visual interaction pattern does have an impact on the dynamics in the absence of social interaction, confirming the results obtained in its presence when participants share direct visual coupling \cite{alderisio2016interaction}.
We also investigate the effects of a virtual player joining a group of $4$ persons on the overall group dynamics, presenting for the first time in the literature some preliminary experimental results on the effect that virtual agents can have on the emergence of coordination in a human ensemble. All the experiments reported in the paper serve to illustrate the effectiveness and flexibility of the new platform we present, which is available for download from \url{https://dibernardogroup.github.io/Chronos}.

\section{Materials and methods}
This study was carried out in accordance with the recommendations of the local ethical committees of the University of Naples ``Federico II'' and the University of Bristol. All subjects gave written informed consent in accordance with the Declaration of Helsinki.

\subsection{Computer-based set-up architecture}

The proposed computer-based platform is a hardware/software set-up consisting of input/output devices, a centralized unit (server and client-adiministrator) processing data, broadcasting movement information to the various client-players and implementing virtual agents, and a Wi-Fi apparatus connecting all the components together. 
The central server unit receives position data from the client-players and broadcasts to each position data from a subset of the others, according to the desired structure of interconnections being implemented. For example, in a ring network each client-player will only receive position data from two neighboring client-players. The movements of each human agent are detected by a low-cost position sensor, and individuals interact with each other through their own personal computer, on whose screens the central unit broadcasts the appropriate position trajectories according to the assigned topology (visual interaction patterns). The central unit is also responsible for data management and storage.

\begin{figure}[ht!]
\begin{center}
\includegraphics[width=13cm]{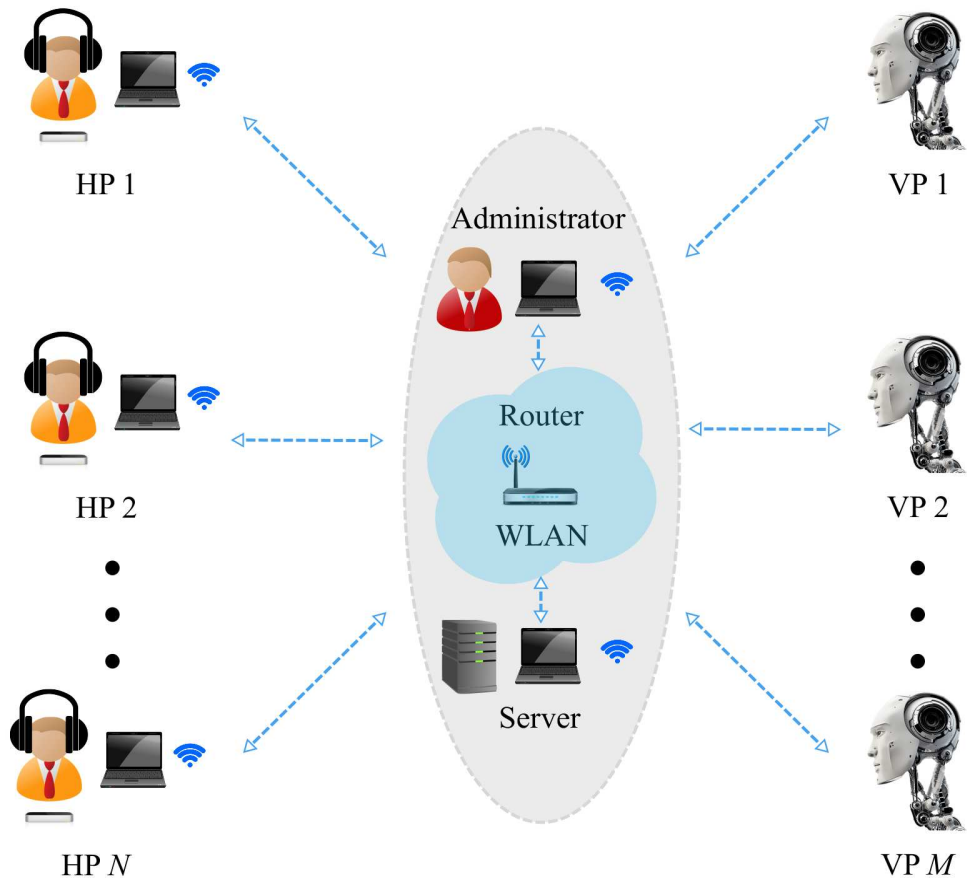}
\end{center}
\caption{\textbf{Computer-based set-up architecture.} $N$ player modules, respectively accessible by one human participant (HP) each, send the Server requests to perform either \emph{Dyadic} or \emph{Group interaction} trials, so that data can be appropriately stored and players can interact with each other in real-time. $N$ human players, hearing white noise through headphone sets, move their preferred hand over their own motion sensor. They see their own 1D position trajectory and that of the others they are possibly interacting with on their respective computer screen. An Administrator module allows to record (and store through the Server in an appropriate database) the motor signature of a given human participant in \emph{Solo experiments}, and to set the topology of interconnections, the duration of each trial and the model of $M$ possible virtual players (VP) in the case of \emph{Dyadic} and \emph{Group interaction}. A Server module implements the virtual players as computerized versions of the human motion, without the need for additional machines or physical entities/robots. All the machines are connected onto the same wireless local area network (WLAN) by means of a dedicated Wi-Fi router.}\label{fig:1}
\end{figure}

The proposed set-up is shown in details in Figure \ref{fig:1} for the case of $N$ human participants and $M$ virtual agents, and described below in all its components (for more information on how to use the set-up and for download of the software, see \url{https://dibernardogroup.github.io/Chronos} and Section \ref{sec:suppInf3} of Supplementary Material).

\subsubsection{Hardware equipment}
\label{sec:he}
\begin{figure}[h!]
\begin{center}
\includegraphics[width=15cm]{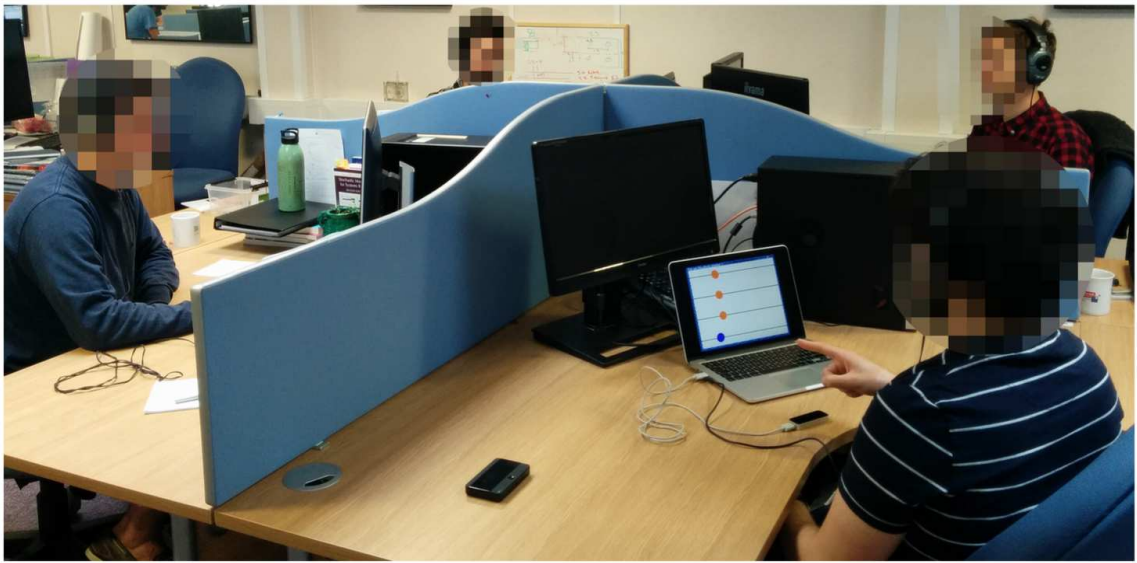}
\end{center}
\caption{ \textbf{Experimental set-up in \emph{Group interaction} experiments.} Human participants move their preferred hand over a Leap Motion controller while sitting around a table. They are separated by barriers (no direct visual coupling) and wear headphones (no auditory coupling), so that social interaction is removed. For each participant, a blue circle on the screen represents her/his own motion, while orange circles represent those of the others s/he is possibly coupled with.}\label{fig:2}
\end{figure}

The hardware equipment consists of:
\begin{itemize}
\item $N$ \emph{low-cost motion sensors}. Each player waves the index finger of her/his preferred hand over a Leap Motion controller (Leap Motion, Inc.) which captures its movements over time as a monodimensional trajectory \cite{Guna2014analysis}; alternatively, a mouse or trackpad can be used.
\item $N+1$ \emph{Personal Computers}. Each motion sensor is connected to a PC, such that the recorded position trajectory can be stored after any trial. Participants are able to see their motion and that of the others they are possibly interacting with on their respective computer screens, by means of moving color-coded circles. For each participant, a blue circle represents her/his own motion, whereas orange circles represent those of the others, respectively (see Supplementary Figure \ref{fig:s6} of Supplementary Material). The two coordinates of each circle on the screen are updated according to the position detected by the motion sensor: one of them is kept fixed, while the other corresponds to the input received by the motion sensor. One additional computer is needed to run the server and a GUI that allows the administrator to set the experimental parameters and the desired visual interaction patterns (see Section \ref{sec:ss}). No further machines are needed to implement the $M$ virtual players, as the cognitive architecture driving their motion is run by the central server that dispatches their position data to the various clients as required.
\item $N$ \emph{headphone sets}. Each player wears headphones through which white noise is transmitted to eliminate possible auditory couplings with the others.
\item $1$ \emph{router}. It provides Wi-Fi signal in order to allow clients (administrator and players) and server to be logged onto the same wireless local area network (WLAN) through TCP/IP protocol \cite{forouzan2002tcp}.
\end{itemize}
Furthermore, barriers are employed to separate the players and prevent them from being directly visually coupled (Figure \ref{fig:2}).

\subsubsection{Software architecture}
\label{sec:ss}
The software architecture, which is based on a client-server model \cite{berson1996client}, consists of:
\begin{itemize}
\item $N$ \emph{Player modules}. These modules, respectively accessible by one human player each, provide a user-friendly interface through which participants interact. They send the server requests to perform either \emph{Dyadic} or \emph{Group interaction} trials (see Section \ref{sec:toe}), such that data can be appropriately stored and the trial correctly started.
\item $1$ \emph{Administrator module}. This module, accessible by the administrator only, carries out different tasks according to the type of experiments being performed (see Section \ref{sec:toe}) through a user-friendly interface. In the case of \emph{Solo experiments}, it allows to record the motor signature of a given human participant. In the case of \emph{Dyadic interaction}, it allows to set the duration of each trial, the roles played by the two agents, as well as model and parameter of the possible VP to be used in human-virtual player experiments. In the case of \emph{Group interaction}, it allows to set the topology of interactions among the HPs and the duration of each trial, as well as to choose the number of VPs and their models and parameters to be used in mixed human-virtual player experiments.
\item $1$ \emph{Server module}. It handles communication among different players' machines and connects them onto the WLAN provided by the router. In the case of \emph{Solo experiments}, it deals with the motor signature storage in an appropriate database. In the case of \emph{Dyadic} and \emph{Group interaction}, it manages requests coming from the players so that all the participants can interact in real-time, dealing with trajectories broadcasting (each player sees her/his motion and that of the others s/he is interacting with) and storage.
\end{itemize}

\subsection{Types of experiments}
\label{sec:toe}
\begin{figure}[h!]
\begin{center}
\includegraphics[width=17cm]{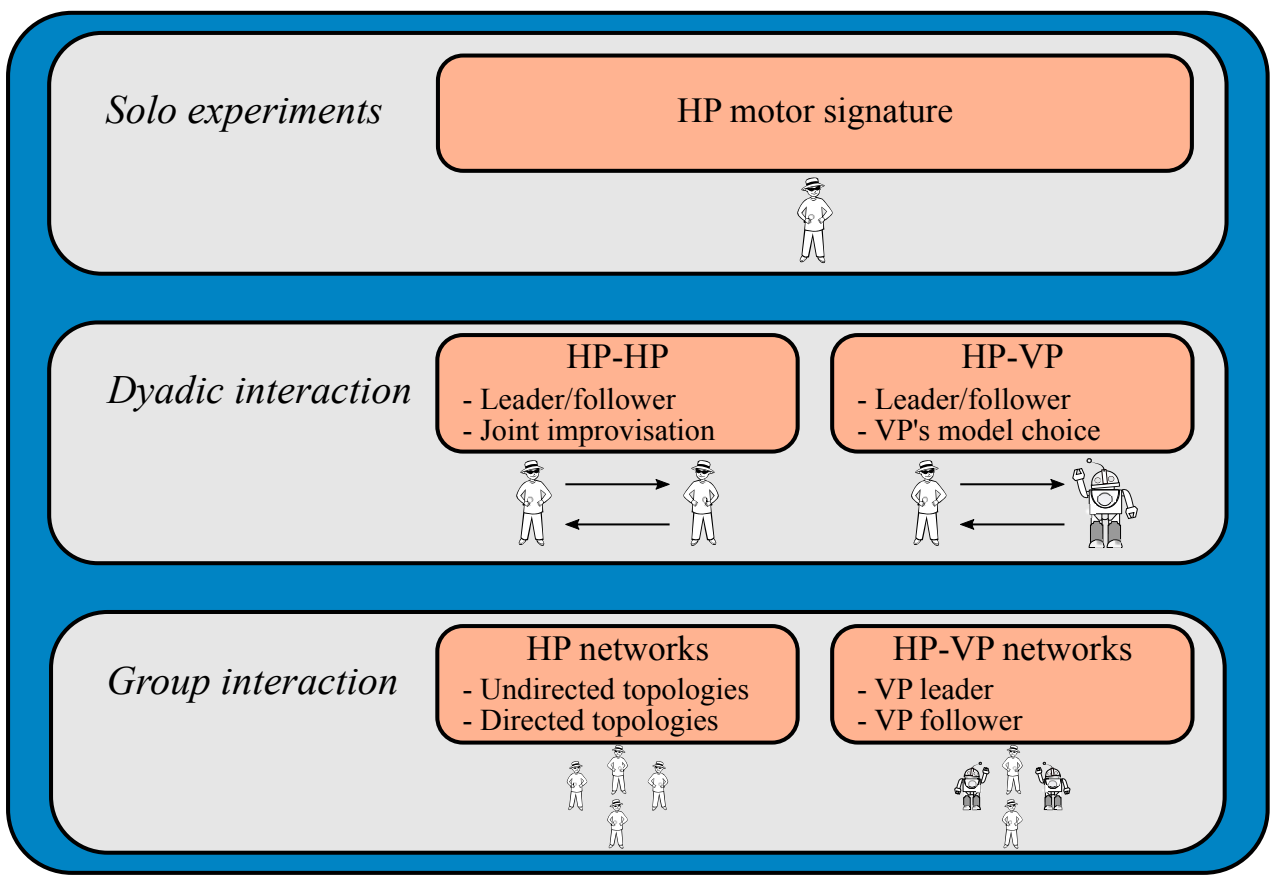}
\end{center}
\caption{ \textbf{Choice of experiments through our proposed technology.} \emph{Solo experiments}: participants are separately asked to generate a spontaneous movement of their preferred hand in isolation, so that their individual motor signature can be recorded. \emph{Dyadic interaction}: both HP-HP (two human participants can either interact in a Leader-Follower or in a Joint improvisation condition) and HP-VP trials (a human participant is asked to either lead or follow a virtual agent, whose mathematical description for its dynamics can be chosen among different models) can be performed. \emph{Group interaction}: any kind of structure of interconnections can be set among the players. Three or more human participants are asked to synchronize the motion of their preferred hand with that of the others they are topologically connected to (HP networks), with the possibility of implementing virtual agents in the group (HP-VP networks), which can be set to act either as follower or leader.}\label{fig:3}
\end{figure}

All types of experiments that can be performed through the proposed technology are listed below and summarized in Figure \ref{fig:3}.

\begin{enumerate}
\item \emph{Solo experiments}. These experiments involve only one agent at a time. Participants are separately asked to generate some spontaneous movement of their preferred hand, so that their individual motor signature as defined in \cite{slowinski2016dynamic} can be acquired.
\item \emph{Dyadic interaction}. These experiments involve only two agents. Two kinds of trials can be performed:
\begin{itemize}
\item HP-HP trials: human participants can either interact in a Leader-Follower condition (one of them leads the game and the other tracks her/his hand movements), or in a Joint Improvisation condition (there is no designation of leader and follower, the two participants are asked to create an interesting and synchronized motion of their preferred hands).
\item HP-VP trials: a human participant is asked to either lead or follow a virtual agent, whose mathematical description for its dynamics can be chosen among different models (see Section \ref{sec:suppInf1} of Supplementary Material and \cite{zhai2014adaptive,zhai2014novel,zhai2015design,alderisio2016comparing} for further details).
\end{itemize}
\item \emph{Group interaction}. These experiments involve three or more agents, where any kind of structure of interconnections among them can be set. In particular, the network topology can be either \emph{undirected} (participant $i$ sees the motion of participant $j$ if and only if participant $j$ sees the motion of participant $i$) or \emph{directed} (the previous condition is not verified). Two kinds of networks can be implemented:
\begin{itemize}
\item HP networks: three or more human participants are asked to synchronize the motion of their preferred hand with that of the others they are topologically connected with.
\item mixed HP-VP networks: one or more participants of the group are virtual agents, which can be set to act either as followers or leaders, according to how much attention they pay to tracking the motion of the other group members they are connected with or generating spontaneous movements, respectively (see Section \ref{sec:suppInf1} of Supplementary Material and \cite{zhai2014adaptive,zhai2014novel,zhai2015design,alderisio2016comparing} for further details).
\end{itemize}
\end{enumerate}

\subsection{Experimental set-up validation}

\subsubsection{Participants}
A total of $9$ people participated in the experiments: $1$ female and $8$ males (all the participants were right handed). The participants, who volunteered to take part in the experiments, were master and PhD students from University of Naples Federico II in Italy, and PhD students and Postdoctoral Research associates from University of Bristol in the UK. The experiments took place in three separate sessions.

\subsubsection{Task and procedure}
Different tasks were performed by means of the proposed computer-based set-up, where participants were asked to sit around a table and move the index finger of their preferred hand as smoothly as possible over a Leap Motion controller, along a direction required to be straight and parallel to the floor.

\begin{enumerate}
\item \emph{Solo experiments}. Four participants were asked to separately perform $4$ trials, each of duration $60$s. Specifically, each participant was told to perform $2$ trials while producing a sinusoidal-like wave at their own natural oscillation frequency, and then $2$ more trials while producing an interesting non-periodic motion representing their motor signature \cite{slowinski2016dynamic}. These experiments were performed in the first session.
\item \emph{Dyadic interaction}. The same four participants were grouped in two pairs: players $1$ and $2$ formed Dyad 1, while players $3$ and $4$ formed Dyad 2, respectively.
\begin{itemize}
\item Each player was asked to perform $2$ HP-HP trials of duration $30$s in Leader-Follower condition, and did not know the identity of her/his partner. In particular, players $1$ and $4$ acted as leader, while players $2$ and $3$ as followers.
\item Then, for each pair, either of the two players was replaced by a virtual agent (modeled by HKB equation with PD control, see Section \ref{sec:suppInf1} of Supplementary Material and \cite{zhai2014adaptive,zhai2014novel,zhai2015design,alderisio2016comparing} for further details) fed with the motor signature, captured during \emph{Solo experiments}, of the human player it was substituting (Figure \ref{fig:4}). In particular, players $1$ and $3$ were replaced, and players $2$ and $4$ were not informed on this (they believed they were still interacting with their human partner). Once again, $2$ HP-VP trials of duration $30$s were performed for each pair in Leader-Follower condition.
\end{itemize}
These experiments were performed in the second session.

\begin{figure}
\begin{center}
\includegraphics[width=17cm]{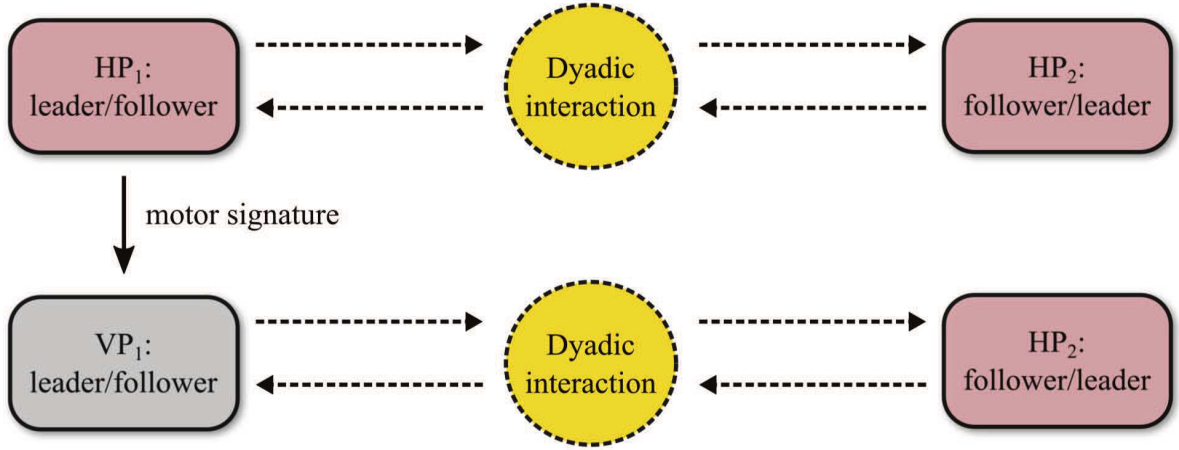}
\end{center}
\caption{ \textbf{\emph{Dyadic interaction} experiments.} Two human participants are asked to perform trials in a Leader-Follower condition. Then one of them is replaced by a virtual player, which is provided with the same kinematic features (motor signature) as those of the substituted human player. The virtual agent plays the role of the replaced human participant in the HP-HP interaction (Leader or Follower).}\label{fig:4}
\end{figure}

\item \emph{Group interaction}. Two different groups of four (the same as \emph{Solo experiments} and \emph{Dyadic interaction}) and five other participants were separately tested, respectively named Group 1 and Group 2. Participants in each group were asked to synchronize their motion with that of the circles shown on their respective computer screen, representing the movements of the other agents topologically connected with them. However, players had no global information of the topology of their interactions.
\begin{itemize}
\item Group 1. Four participants were involved in this session. Firstly, $3$ trials of $30$s each were performed where all participants saw on their respective screens traces of the objects moved by all the others (all-to-all configuration, Figure \ref{fig:5}A). Secondly, a VP (modeled by HKB equation and adaptive control, see Section \ref{sec:suppInf1} of Supplementary Material and \cite{zhai2014adaptive,zhai2014novel,zhai2015design,alderisio2016comparing} for further details) fed with the sinusoidal motion of a different player was introduced in the network; participants were told that a fifth human player was interacting with them. The virtual agent was first connected in \emph{leader mode} to either $1$, $2$ or $4$ HPs (Figures \ref{fig:5}B-D, respectively), and then in \emph{follower mode} to all of them (Figure \ref{fig:5}E). For each topology including the virtual player, once again $3$ trials of duration $30$s were performed. These experiments were performed in the second session.

\begin{figure}[h!]
\begin{center}
\includegraphics[width=17cm]{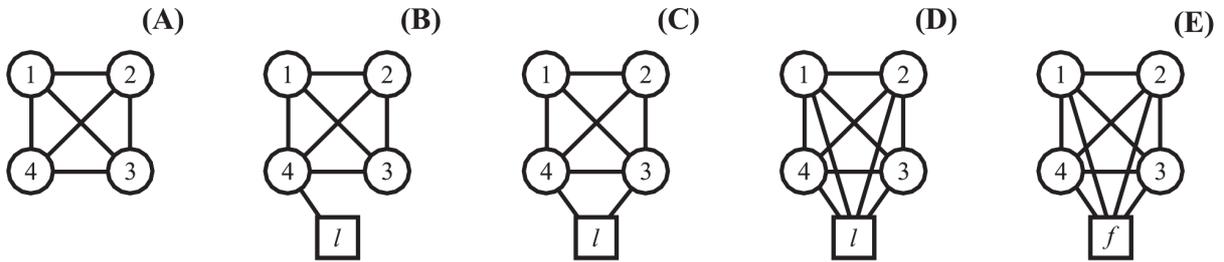}
\end{center}
\caption{ \textbf{Topology of connections among participants in the \emph{Group interaction} experiments -- Group 1.} Circles refer to human participants, while the square refers to the virtual player (\emph{l}: leader mode, \emph{f} : follower mode). (A) Undirected all-to-all interaction structure (each player sees the motion of all the others), with the addition of undirected links between a VP and $1$, $2$ or $4$ HPs (B-E, respectively) are shown. The virtual player acts as a leader in topologies (B-D) and as a follower in topology (E).}\label{fig:5}
\end{figure}

\item Group 2. Five participants were involved in this session. Eight different topologies of interactions were implemented among them (Figure \ref{fig:6}): undirected complete (Figure \ref{fig:6}A), ring (Figure \ref{fig:6}B), path (Figure \ref{fig:6}C) and star graph (Figure \ref{fig:6}D), and their respective directed version (Figures \ref{fig:6}E-H).

\begin{figure}[h!]
\begin{center}
\includegraphics[width=17cm]{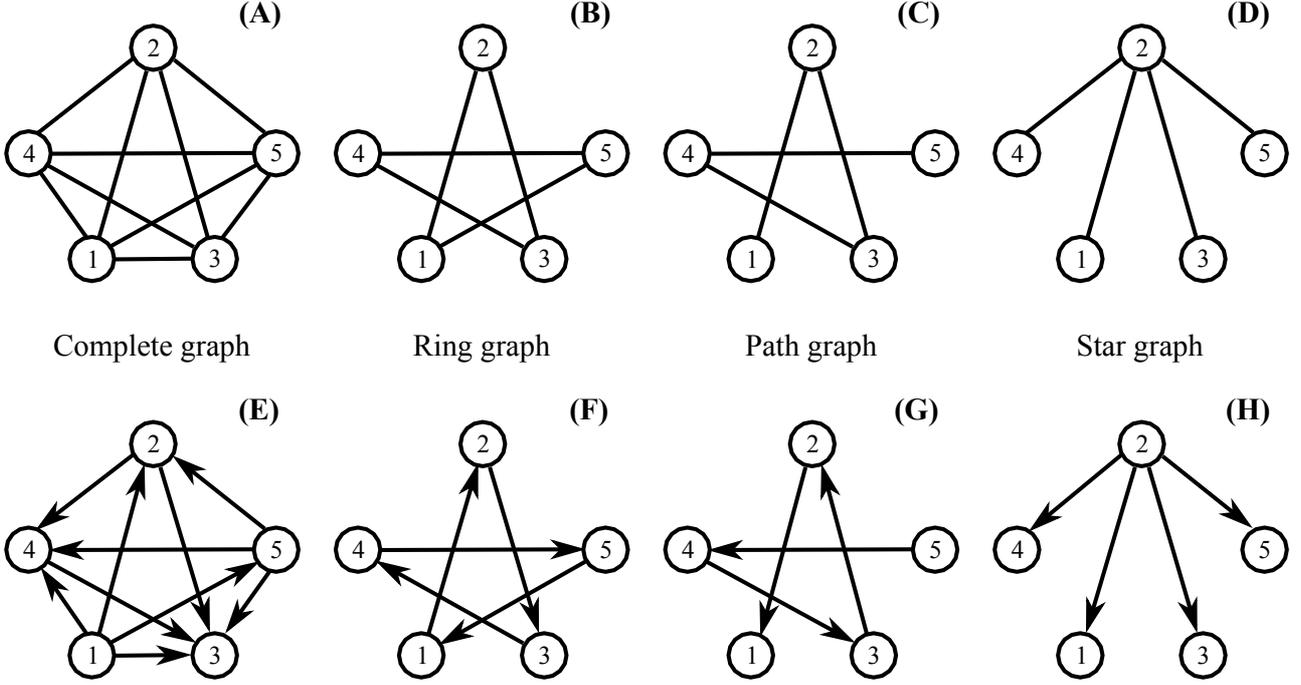}
\end{center}
\caption{ \textbf{Topology of connections among participants in the \emph{Group interaction} experiments -- Group 2.} (A-D) represent undirected complete, ring, path and star graph, respectively. (E-H) represent the respective directed versions. Edges without arrows represent \emph{undirected} connections (if participants $i$ sees the motion of participant $j$, then also participant $j$ sees the motion of participant $i$), whereas in the \emph{directed} case, an edge going out of node $i$ and coming in node $j$ (the direction of the edge is given by its corresponding arrow) is representative of the fact that participant $j$ sees the motion of participant $i$.}\label{fig:6}
\end{figure}

As for the undirected topologies:
\begin{itemize}
\item[-] Complete graph: each participant could see the movements of all the others.
\item[-] Ring graph: each participant could see the movement of only two other players, called \emph{neighbors}.
\item[-] Path graph: similar to the ring graph configuration, but two participants (players $1$ and $5$), defined as \emph{external} participants, could see the movements of only one \emph{neighbor} (respectively players $2$ and $4$), and as a consequence were not connected to each other.
\item[-] Star graph: one participant defined as \emph{central} (player $2$) could see the movements of all the others, defined as \emph{peripheral}, who in turn could see the movements of only the \emph{central} player.
\end{itemize}
For each topology, $6$ trials of duration $30$s were performed. These experiments were performed in the third session.

\end{itemize}
\end{enumerate}

\subsubsection{Data acquisition and analysis}
Since the movements of the participants were mainly monodimensional, only one coordinate of their position trajectory was captured by the Leap Motion. Data was originally stored with a frequency rate of $10Hz$ ($13Hz$ for Group 2), and then underwent cubic interpolation ($100Hz$, see Supplementary Figure \ref{fig:s1} of Supplementary Material).

\subsubsection{Synchronization metrics}

Next we briefly describe all the metrics used to assess players' performance in \emph{Dyadic} and \emph{Group interaction} experiments (refer to Section \ref{sec:suppInf2} of Supplementary Material for further details and mathematical definitions of all metrics introduced below).

In \emph{Dyadic interaction experiments}, the relative phase $\phi_{d_{h,k}} \in [-\pi,\pi]$ was used to check whether the assigned roles of leader and follower were respected during the interaction by participants $h$ and $k$. As $\phi_{d_{h,k}}$ is defined as the difference between the phase of the leader and that of the follower, positive values indicate that the designated leader is effectively leading the game while interacting with the follower \cite{zhai2016design}.
In addition, the symmetric \emph{dyadic synchronization index} $\rho_{d_{h,k}} \in [0,1]$ was used to quantify the average coordination level between agents $h$ and $k$ over time (the closer $\rho_{d_{h,k}}=\rho_{d_{k,h}}$ is to 1, the lower the phase mismatch within the pair $(h,k)$ over the whole trial duration). The root mean square (RMS) of the normalized position error $\epsilon_{h,k} \in[0,100] \%$ was also employed as a measure of the position mismatch (expressed in percentage) between the two agents.

In \emph{Group interaction experiments}, the coordination level among all players was quantified by means of the \emph{group synchronization index} $\rho_g(t) \in [0,1]$, which represents the coordination level of the group at time $t$ (the closer $\rho_g(t)$ is to 1, the lower the average phase mismatch of all the agents in the ensemble at time $t$). Its mean value over time $\rho_g$ was used as a measure of the coordination level of the group over the whole trial.

Note that group and dyadic synchronization indices, originally introduced in \cite{richardson2012measuring}, were evaluated from the phases of the players' movements, which in turn were estimated by making use of the Hilbert transform \cite{kralemann2008phase}.

\section{Results}
\label{sec:res}

\subsection{Virtual players can successfully interact with humans}
\begin{figure}[h!]
\begin{center}
\includegraphics[width=17cm]{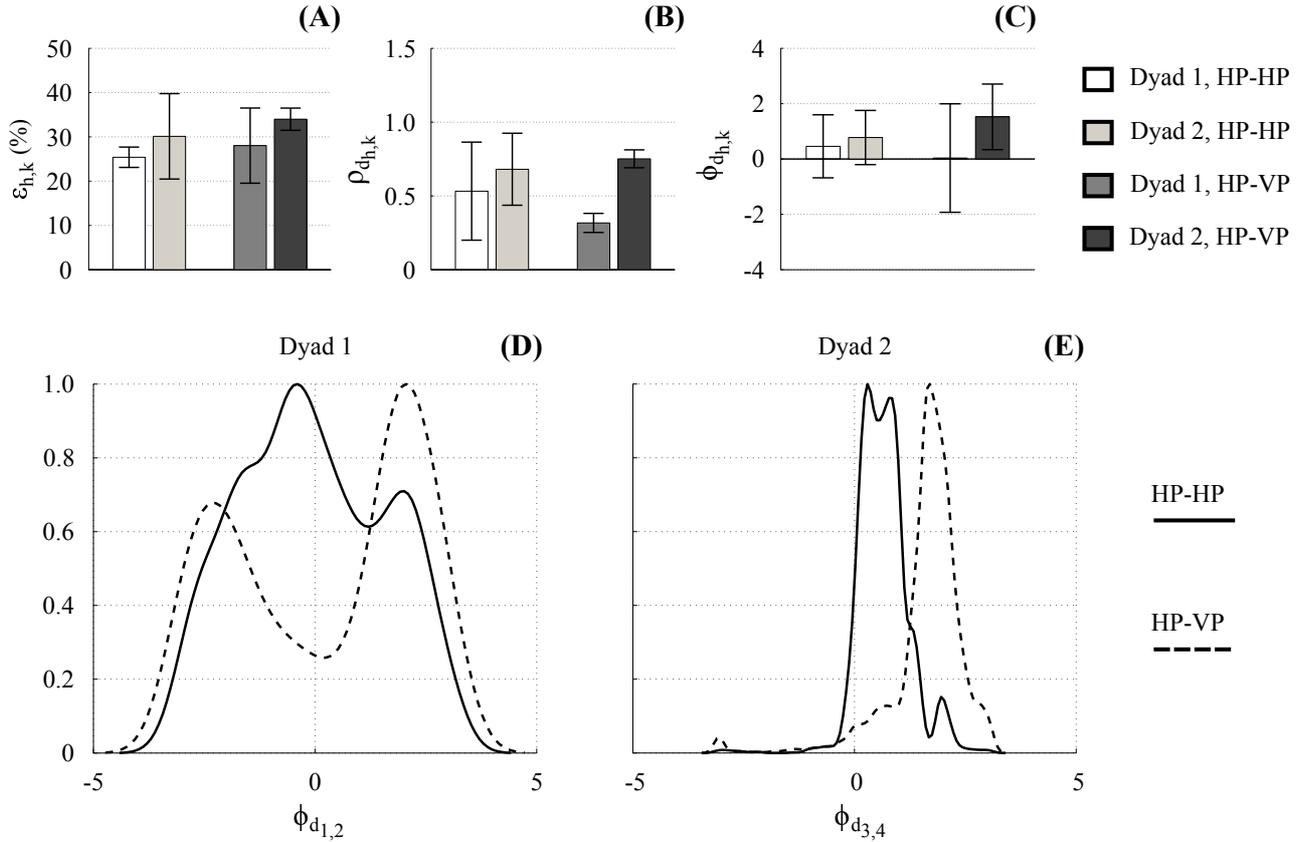}
\end{center}
\caption{ \textbf{Experimental results in the \emph{Dyadic interaction} experiments.} RMS of the normalized position error $\epsilon_{h,k}$ (A), dyadic synchronization indices $\rho_{d_{h,k}}$ (B) and relative phase $\phi_{d_{h,k}}$ between the two participants (C) are shown for each pair (Dyad 1 and Dyad 2), where different scales of grey refer to different pairs and players. The height of each bar represents the mean value averaged over the $3$ trials for each pair, whereas the black error bar represents its averaged standard deviation. The PDF of the relative phase $\phi_{d_{h,k}}$ between the two participants of Dyad 1 (D) and Dyad 2 (E) are shown for the first trial of each pair, where the black solid line refers to HP-HP interaction, and the black dashed line refers to HP-VP interaction.}\label{fig:7}
\end{figure}

The relationships between the metrics obtained for the two dyads in HP-HP interaction are replicated when substituting one of the two human players in each pair with a virtual agent (Figure \ref{fig:7}).
This confirms that the VP, as designed in \cite{alderisio2016comparing} and implemented in our novel software set-up, is able to interact in a human-like fashion with the other player, becoming a kinematic avatar of the person it is substituting in the game \cite{zhai2016design}.
In particular, the RMS of the normalized position error $\epsilon_{1,2}$ obtained in Dyad 1 is lower than $\epsilon_{3,4}$ obtained in Dyad 2 (Figure \ref{fig:7}A), and the same applies to the dyadic synchronization indices $\rho_{d_{1,2}}$ and $\rho_{d_{3,4}}$ (Figure \ref{fig:7}B), and for the relative phase $\phi_{d_{1,2}}$ and $\phi_{d_{3,4}}$ (Figure \ref{fig:7}C). Notably, for both dyads, the probability density function (PDF) of the relative phase obtained for the two players in HP-VP interaction resembles that obtained in HP-HP interaction (Figures \ref{fig:7}D-E). Indeed, the PDFs related to Dyad 1 are broader ad centred around $0$, whereas those related to Dyad 2 are tighter and shifted on the right.

\subsection{Effects of visual interaction patterns on group coordination}
\begin{figure}[!ht]
\begin{center}
\includegraphics[width=15cm]{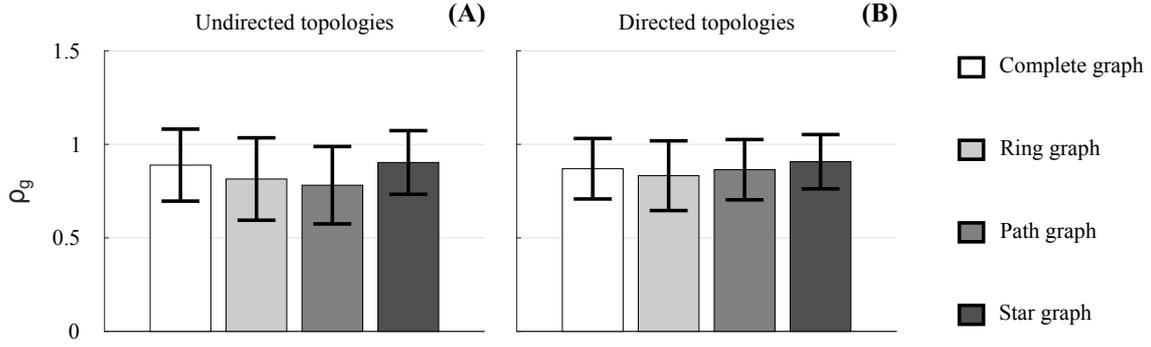}
\end{center}
\caption{ \textbf{Coordination level in the \emph{Group interaction} experiments -- Group 2.} Different scales of grey refer to different topologies. The height of each bar represents the mean value over time of the group synchronization index $\rho_g(t)$, averaged over the total number of trials, whereas the black error bar represents its averaged standard deviation. The coordination levels in (A) undirected and (B) directed topologies are shown.}\label{fig:8}
\end{figure}

The group synchronization observed experimentally in Group 2 depends on the topology of the interconnections among the participants, with complete graph and star graph exhibiting the highest values both in the undirected and in the directed topologies (Figure \ref{fig:8}).
This result confirms independently the observations reported in \cite{alderisio2016interaction} showing that interaction patterns among participants have a significant effect on their coordination level.
Also, for each topology, higher mean values and lower standard deviations are observed in the directed case, with the only exception of the former in the complete graph (for more details see Supplementary Table \ref{tab:3} of Supplementary Material).

As expected for both undirected and directed topologies, in most cases ($83\%$ for undirected and $91\%$ for directed topologies) the highest mean values of dyadic synchronizations over the total number of trials are observed within topologically connected participants (Figure \ref{fig:9}). For more details see Supplementary Tables \ref{tab:4} and \ref{tab:5} of Supplementary Material.

\begin{figure}[!hb]
\begin{center}
\includegraphics[width=17cm]{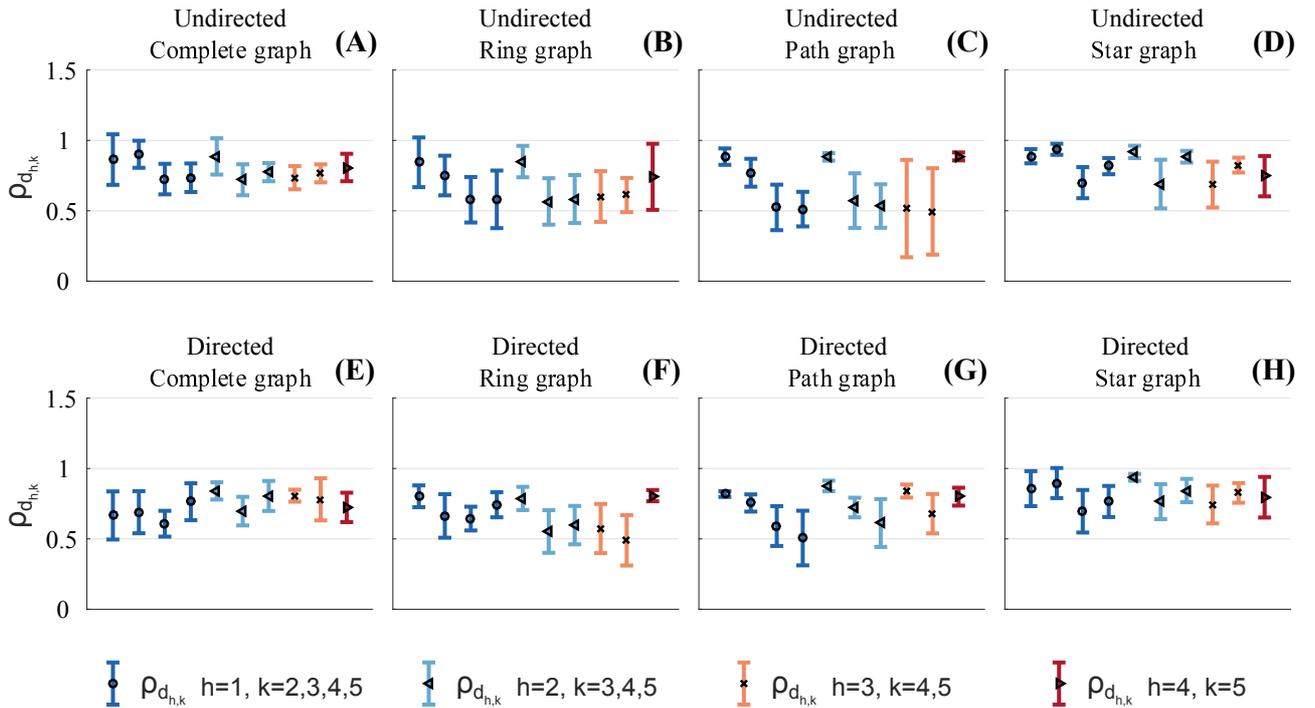}
\end{center}
\caption{ \textbf{Dyadic synchronization index in the \emph{Group interaction} experiments -- Group 2.} Different symbols and colors refer to pairs related to different players. Mean (symbol) and standard deviation (error bar) over the total number of trials of the dyadic synchronization indices $\rho_{d_{h,k}}=\rho_{d_{k,h}}$ in the undirected ((A) complete, (B) ring, (C) path and (D) star graph) and in the respective directed topologies (E-H) are shown. As $\rho_{d_{h,k}}$ are symmetric by definition, only half of them are depicted.}\label{fig:9}
\end{figure}

\subsection{Effects of virtual players on group coordination}
\begin{figure}[t]
\begin{center}
\includegraphics[width=17cm]{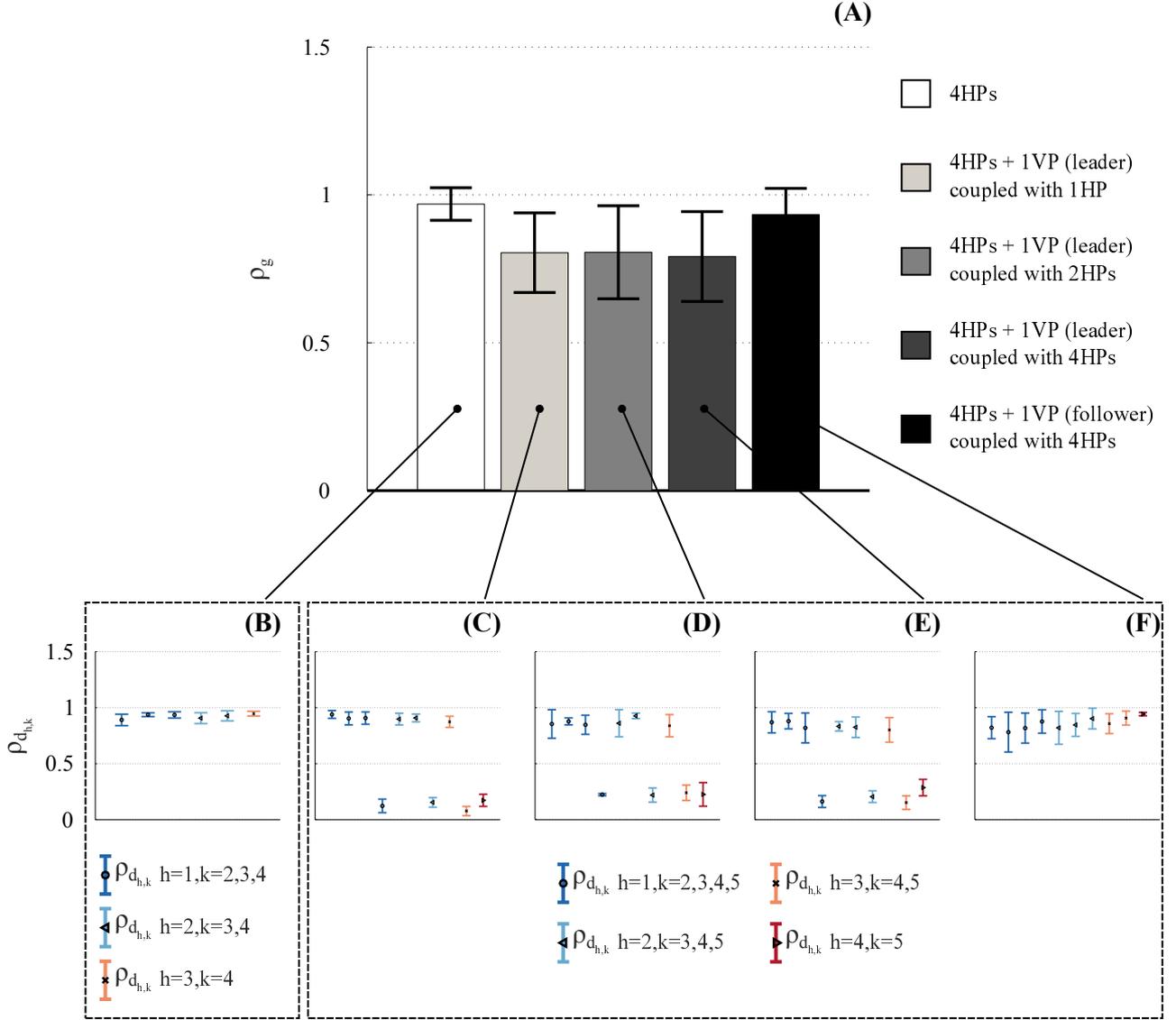}
\end{center}
\caption{ \textbf{Experimental results in the \emph{Group interaction} experiments -- Group 1.} The group synchronization indices obtained for the players in the five different topologies of Figure \ref{fig:5} are shown, with different scales of grey representing different topologies (A). The height of each bar represents the mean value over time of the group synchronization index $\rho_g(t)$, averaged over the $3$ trials for each topology, whereas the black error bar represents its averaged standard deviation. The corresponding dyadic synchronization indices $\rho_{d_{h,k}}$ obtained for all the pairs of players in the topologies of Figures \ref{fig:5}A-E are respectively shown in panels (B-F). Different symbols and colors refer to mean and standard deviation averaged over the $3$ trials performed for each topology, respectively. As $\rho_{d_{h,k}}$ are symmetric by definition, only half of them are depicted.}\label{fig:10}
\end{figure}

Using the ability of the platform presented in this paper to deploy virtual players during \emph{Group interaction} experiments, we evaluate next the effect of introducing a VP, either as a leader or a follower, in a group of HPs performing a joint task. We observe that the highest value of group synchronization observed experimentally is obtained in the HP network, while lower values are obtained when introducing a VP as leader. However, the group synchronization index $\rho_g$ increases again when a VP is introduced as follower (Figure \ref{fig:10}A). These results are confirmed by the dyadic synchronization indices $\rho_{d_{h,k}}$  respectively obtained in the five topologies of interest (Figures \ref{fig:10}B-F). For each pair of human players, high values (Figure \ref{fig:10}B) are observed for the topology shown in Figure \ref{fig:5}A. On the other hand, when a virtual leader is introduced in the interaction (topologies shown in Figures \ref{fig:5}B-D) the lowest values of dyadic synchronization are obtained for each human player in correspondence to the VP (player $5$ in Figures \ref{fig:10}C-E). Finally, when the VP acts as a follower (topology shown in Figure \ref{fig:5}E), the highest values of dyadic synchronization indexes $\rho_{d_{h,k}}$ for each human player are observed in correspondence to the VP (player $5$ in Figure \ref{fig:10}F). For more details see Supplementary Tables \ref{tab:1} and \ref{tab:2} of Supplementary Material.

\section{Discussion}

In this work we investigated human coordination, both in dyads and in groups over different structures of interconnections, proposing an \emph{ad hoc} novel computer-based set-up that allows to remove the effects of social interactions among the players and deploy virtual agents in the group, thus opening the possibility of further investigating the mechanisms that underly human group coordination through an extension of the \emph{human dynamic clamp} to multiplayer scenarios \cite{dumas2014human}.

We envisage that the computer set-up presented in this paper can be used in Social Psychology to investigate what the effects of social interactions are in dyadic or group movement coordination. Indeed, joint action tasks might first be performed while allowing participants to share direct visual and auditory coupling (participants directly look at each other instead of the screen of their personal computers, and do not wear headphones so that they know who they are interacting with), and then while removing them (or vice versa).
Moreover, since some of the players can be replaced with one or more virtual agents, our computer technology can also be exploited for the development of artificial agents able to merge and interact within a group of humans \cite{boucenna2016robots,iqbal2016movement}, both for recreational \cite{alac2011robot} and rehabilitation purposes \cite{zhai2015model,bono2016goliah}.

In order to validate our experimental set-up, we applied it to test two significant claims recently made in the literature, and explored the effect of deploying a virtual player during group coordination experiments.
Specifically, we validated the use of a virtual player as designed in \cite{alderisio2016comparing} in a dyadic interaction task, and confirmed the results of \cite{zhai2016design} showing that virtual players can successfully interact with humans. We found that the behavior exhibited in terms of the metrics used in Section \ref{sec:res} by each dyad was the same for both HP-HP and HP-VP interaction, meaning that human players did not change their way of interacting with their partner according to the nature of the latter. 
In particular, we observed that if a human participant, to whom a follower role was assigned in duo interaction, ended up also leading her/his human partner (despite the instruction given), s/he did so also when interacting with a virtual leader (Dyad 1). On the other hand, if a human leader was successfully leading her/his human partner, s/he did so also in the interaction with a virtual follower (Dyad 2).
Moreover, we confirmed that group coordination level in a human ensemble is affected by the specific structure of interconnections among group members when any form of direct visual, auditory or social interaction is removed, a result found also in \cite{alderisio2016interaction} yet in the presence of visual and social cues. Also, we observed that virtual players can decrease group coordination levels when acting as leaders. Higher values of group and dyadic synchronization indices were instead observed either when no virtual player was interacting within the human ensemble, or when it was following the motion of all the subjects.
This confirms that virtual players can be used to vary the level of coordination in a human group, although further work is required to better understand this effect and its implications.

Some further extensions to our work include the possibility of implementing time-varying topologies to study the effects of adding/removing connections among interacting participants \cite{cardillo2014evolutionary}, and enabling the administrator to provide the players with social cues in real time, based on the quality of their performance (i.e., as measured by the group synchronization index). Finally, it is possible to implement new mathematical models, based on optimal control theory, for the VP to perform as joint improviser \cite{zhai2016design} with other virtual or human agents.

\clearpage

\section*{Conflict of Interest Statement}
The authors declare that the research was conducted in the absence of any commercial or financial relationships that could be construed as a potential conflict of interest.

\section*{Author Contributions}
Conceived and designed the experiments: FA, GF, MdB. Performed the experiments: FA, ML, GF. Analyzed the data: FA, GF. Contributed analysis tools: FA, GF. Developed the software: ML. Wrote the paper: FA, GF, MdB.

\section*{Funding}
This work was supported by the European Project AlterEgo FP7 ICT 2.9 - Cognitive Sciences and Robotics, Grant Number 600610.

\section*{Acknowledgments}
We wish to thank all the people taking part in the experiments. 

\section*{Supplementary Material}
Supplementary material accompanies this paper.

\clearpage

\section*{SUPPLEMENTARY MATERIAL}

\section{Mathematical model of the virtual player}
\label{sec:suppInf1}
Let $x \in \mathbb{R}$ be the state variable representing the position of the virtual player (VP). The system describing its behavior is given by the following dynamical system:

\begin{equation}
\ddot{x}(t)=f(x(t),\dot{x}(t))+u(t)
\end{equation}
where $f$ represents the vector field modelling the inner dynamics of the VP when disconnected from any other agent, $\dot{x}$ and $\ddot{x}$ represent velocity and acceleration of the VP, and $u$ is the control signal that models how the VP interacts with other players.

Our computer-based set-up allows the selection of different combinations of inner dynamics models and control signals to describe the motion of the virtual player (see also Section \ref{sec:su} below). These can be listed as follows.
\begin{itemize}
\item Alternative models for the inner dynamics:
\begin{enumerate}
\item \emph{Harmonic oscillator}, a linear system given by
\begin{equation}
f(x,\dot{x}) = -(a\dot{x}+bx)
\end{equation}
where $a$ and $b$ represent viscous damping coefficient and the elastic coefficient, respectively.
\item \emph{HKB equation}, a nonlinear oscillator given by
\begin{equation}
f(x,\dot{x}) = -(\alpha x^2 + \beta \dot{x}^2-\gamma)\dot{x} - \omega^2 x
\end{equation}
where $\alpha, \beta, \gamma$ characterize the damping coefficient, while $\omega$ is related to the oscillation frequency, respectively.
\end{enumerate}
In our computer-based set-up, we also give the opportunity to describe the behavior of the VP as a \emph{double integrator}, that is a system without any inner dynamics whose motion is entirely determined by the coupling with the other agents via the control input $u(t)$. In this case the equation becomes:
\begin{equation}
\ddot{x}(t)=u(t)
\end{equation}
\item Different options for the coupling function $u(t)$:
\begin{enumerate}
\item \emph{PD control}, a linear control law given by
\begin{equation}
\label{eqn:pd}
u = K_p (y-x) + K_\sigma(\dot{\sigma}-\dot{x})
\end{equation}
where $y$ is the position of the other agent coupled to the VP, $\dot{\sigma}$ is its desired motor signature (velocity trajectory), and $K_p$ and $K_\sigma$ are two control gains. According to the values assigned to $K_p$ and $K_\sigma$, the VP acts as a leader (more weight given to $K_\sigma$ so that the VP priority is to minimize the mismatch between its own velocity and that of the prerecorded motor signature) or as a follower (more weight given to $K_p$ and hence higher priority to reducing the mismatch between the VP position and that of the other player).
\item \emph{Adaptive control}, a nonlinear control law which changes according to the nature of the experiment being carried out.
\begin{itemize}
\item[-] When the VP acts as a follower, it is given by
\begin{equation}
u=[\psi+\chi(x-y)^2](\dot{x}-\dot{y})-Ce^{-\delta(\dot{x}-\dot{y})^2}(x-y)
\end{equation}
with
\begin{equation}
\dot{\psi} = -\frac{1}{\psi} [(x-y)(\dot{x}-\dot{y})+(x-y)^2]
\end{equation}
\begin{equation}
\dot{\chi} = -\frac{1}{\chi} (\dot{x}-\dot{y})[f(x,\dot{x})+u]
\end{equation}
where $y$ and $\dot{y}$ are position and velocity of the other agent coupled to the VP, $C$ and $\delta$ are control parameters, and $\psi$ and $\chi$ are adaptive parameters. Note that in this case no motor signature can be assigned to the VP.
\item[-] When the VP acts as a leader, the control input is set as
\begin{equation}
\label{eqn:adaptive}
u=\lambda \left( [\psi+\chi(x-\sigma)^2](\dot{x}-\dot{\sigma})-Ce^{-\delta(\dot{x}-\dot{\sigma})^2}(x-\sigma) \right) + (1-\lambda)K(y-x)
\end{equation}
where $\lambda:=e^{-\delta|x-y|}$, $K$ is a control parameter, $\sigma$ and $\dot{\sigma}$ are desired position and velocity profiles (motor signature) that allow the VP to generate spontaneous motion, and all the other quantities have been previously defined.
\end{itemize}
\end{enumerate}
Note that when the VP is influenced by the motion of two or more agents, as it might happen in the \emph{Group interaction} trials, then $y$ and $\dot{y}$ are appropriately replaced by average position and velocity of all the agents connected to the VP, respectively.
\end{itemize}

More details on each of these mathematical models can be found in \cite{alderisio2016comparing,zhai2015design,zhai2016design,zhai2014adaptive,zhai2014novel,zhai2015model}.

\section{Synchronization metrics}
\label{sec:suppInf2}

Let $x_k(t) \in \mathbb{R} \  \forall t \in [0,T]$ be the continuous time series representing the motion of the $k$-th agent's preferred hand, with $k \in \{ 1, 2, \dots, N \}$, where $N$ is the number of individuals and $T$ is the duration of the experiment. Let $x_k[t_i] \in \mathbb{R}$, with $k \in \{ 1, 2, \dots, N \}$ and $i \in \{ 1, 2, \dots, N_T \}$, be the discrete time series of the position of the $k$-th agent, obtained after sampling $x_k(t)$ at time instants $t_i$, where $N_T$ is the number of time steps of duration $\Delta T := \frac{T}{N_T}$, that is the sampling period. Let $\theta_k(t) \in [-\pi,\pi]$ be the phase of the $k$-th agent, which can be estimated by making use of the Hilbert transform of the signal $x_k(t)$ as detailed in \cite{kralemann2008phase}.

When $N=2$ (\emph{Dyadic interaction}), if we denote with $\phi_{d_{h,k}}(t):=\theta_h(t)-\theta_{k}(t)$ the relative phase between agents $h$ and $k$ at time $t$, it is possible to define the following parameter
\begin{equation}
\label{eqn:r4}
\rho_{d_{h,k}} := \left | \frac{1}{T} \int_{0}^{T} e^{ j \phi_{d_{h,k}}(t) } \ dt \right | \simeq \left | \frac{1}{N_T} \sum_{i=1}^{N_T} e^{ j \phi_{d_{h,k}}[t_i] } \right | \quad \in [0,1]
\end{equation}
as their \emph{dyadic synchronization index}: the closer $\rho_{d_{h,k}}$ is to $1$, the lower the phase mismatch is between agents $h$ and $k$ over the whole trial. 
Note that in a \emph{Leader-Follower} condition, if  $\phi_{d_{h,k}}$ is defined as the difference between the phase of the leader (player $h$) and that of the follower (player $k$), positive values indicate that the designated leader is successfully leading the interaction with the designated follower.

It is also possible to define the root mean square (RMS) of the normalized position error between two agents as
\begin{equation}
\epsilon_{h,k} := \frac{1}{L} \sqrt{\frac{1}{T}\int_0^T \left( x_h(t)-x_k(t) \right)^2 \ dt} \simeq \frac{1}{L} \sqrt{\frac{1}{N_T}\sum_{i=1}^{N_T} \left( x_h[t_i]-x_k[t_i] \right)^2}
\end{equation}
where $L$ refers to the range of admissible position (e.g., the range of motion detected by the Leap Motion controller): the lower $\epsilon_{h,k}$ is, the lower the position mismatch is between agents $h$ and $k$.

When $N>2$ (\emph{Group interaction}), a further index can be used as a measure of the overall coordination level among agents in the group.
We first define the \emph{cluster phase} or \emph{Kuramoto order parameter}, both in its complex form $q'(t) \in \mathbb{C}$ and in its real form $q(t) \in [-\pi,\pi]$ as

\begin{equation}
q'(t) :=  \frac{1}{N} \sum_{k=1}^{N} e^{  j \theta_k(t)  }, \qquad q(t) := {\rm atan2} \left(\Im(q'(t)),\Re(q'(t))  \right) 
\end{equation}
which can be regarded as the average phase of the group at time $t$. 
Denoting with $\phi_k(t) := \theta_k(t) - q(t)$ the relative phase between the $k$-th participant and the group phase at time $t$, we then define the relative phase between the $k$-th participant and the group averaged over the time interval $[0,T]$, both in its complex form $\bar{\phi}'_k \in \mathbb{C}$ and in its real form $\bar{\phi}_k \in [-\pi,\pi]$ as

\begin{equation}
\bar{\phi}'_k := \frac{1}{T} \int_{0}^{T} e^{ j \phi_k(t) } \ dt \simeq \frac{1}{N_T} \sum_{i=1}^{N_T} e^{  j \phi_k[t_i] },\qquad  \bar{\phi}_k := {\rm atan2} \left( \Im(\bar{\phi}'_k), \Re(\bar{\phi}'_k) \right)
\end{equation} 
In order to quantify the synchronization level of the entire group at time $t$, we define the following parameter

\begin{equation}
\label{eqn:r2}
\rho_g(t) := \frac{1}{N} \left | \sum_{k=1}^{N} e^{ j \left( \phi_k(t)- \bar{\phi}_k \right) } \right | \quad \in [0,1]
\end{equation}
as the \emph{group synchronization index}: the closer $\rho_g(t)$ is to $1$, the smaller the average phase mismatch of the agents in the group is at time $t$. 
The mean synchronzation level of the group during the total duration of the performance can be estimated as:

\begin{equation}
\label{eqn:r3}
\rho_g := \frac{1}{T} \int_{0}^{T} \rho_g(t) \ dt \simeq \frac{1}{N_T} \sum_{i=1}^{N_T} \rho_g[t_i] \quad \in [0,1]
\end{equation}

\section{How to use Chronos}
\label{sec:suppInf3}
\label{sec:su}
Here, we briefly explain how to use Chronos, our proposed computer-based set-up. For further details and to download the software, follow the link \url{https://dibernardogroup.github.io/Chronos}.

The hardware equipment necessary for its use consists of:
\begin{itemize}
\item \emph{Low-cost motion sensors}. Each player waves her/his index finger over a Leap Motion controller (Leap Motion, Inc.) which captures its movements over time; alternatively, a mouse or trackpad can be used.
\item \emph{Personal Computers}. Each motion sensor is connected to a PC, such that the recorded position trajectory can be stored after any trial. Participants are able to see their motion and that of the others they are possibly interacting with on their respective computer screens, by means of moving color-coded circles. One additional computer is needed to run the server and a GUI that allows the administrator to set the experimental parameters and the desired visual interaction patterns.
\item \emph{Headphone sets}. Each player wears headphones through which white noise is transmitted to eliminate possible auditory couplings with the others.
\item \emph{Router}. It provides Wi-Fi signal in order to allow clients (administrator and players) and server to be logged onto the same wireless local area network (WLAN) through TCP/IP protocol.
\end{itemize}

For the sake of simplicity, server and administrator modules are run on the same machine, so that they share the same IP address.
Before carrying out any experiments, the administrator needs to follow the preliminary steps listed below (only once):
\begin{enumerate}
\item turn on the wireless router and connect her/his machine to it;
\item import the motor signature database on her/his machine.
\end{enumerate}

For each trial, human players can select the input device (mouse or Leap Motion) and choose to quit the trial before its end by pressing the \emph{q} key on their keyboard. 

\subsection{Solo experiments}
The administrator:
\begin{enumerate}
\item runs the server module on a terminal window;
\item runs the database (MySql server) through \emph{MySql Workbench};
\item runs the administrator module on a different terminal window, and selects \emph{Add new Signature};
\item at the end of the trial, enters the HP's required details and the kind of motion performed (sinusoidal or free) so that her/his position and velocity trajectories ($\sigma$ and $\dot{\sigma}$ in Supplementary Equations \eqref{eqn:pd} and \eqref{eqn:adaptive}, respectively) can be appropriately stored in the database.
\end{enumerate}

The HP performs the experiment.

\subsection{Dyadic interaction}
Two (or three) machines are necessary to carry out HP-VP (or HP-HP) trials.
\begin{itemize}
\item For HP-VP trials, the administrator:
\begin{enumerate}
\item runs the server module on a terminal window of a machine;
\item runs the database (MySql server) through \emph{MySql Workbench};
\item runs the administrator module on a different terminal window and, after selecting \emph{Set dyadic interaction}, selects \emph{HP-VP trials};
\item selects trial duration, roles for the interaction (leader or follower) for both HP and VP, and inner dynamics, control signal and motor signature (extracted from the database) for the VP -- if no parameters are specified, the default values shown in the interface are used.
\end{enumerate}
Note that motor signatures are selected by choosing them among those stored in the database of all signals acquired during previous \emph{Solo experiments}, thus allowing the VP to exhibit the desired kinematic features of one of the HPs whose signatures have been stored during previous experimental runs. Signatures are indexed in the database through a string identifying the players for which they were stored.

\item For HP-HP trials, the administrator:
\begin{enumerate}
\item runs the server module on a terminal window of a machine;
\item runs the administrator module on a different terminal window and, after selecting \emph{Set dyadic interaction}, selects \emph{HP-HP trials};
\item selects the roles for the interaction (leader, follower or joint improviser) for both HPs.
\end{enumerate}
\end{itemize}

For both types of trials, each HP:
\begin{enumerate}
\item connects her/his machine to the WLAN provided by the router;
\item runs her/his client module on a terminal window of her/his machine and, after entering the server's IP address, selects \emph{Dyadic interaction} (Supplementary Figure \ref{fig:s2}) and enters an integer index ($1$ or $2$, respectively) uniquely identifying herself/himself;
\item performs the experiment (her/his role chosen for the interaction by the administrator is shown in the gaming screen).
\end{enumerate}

An example of gaming screen shown to the HP in the case of \emph{Dyadic interaction} experiments (Leader-Follower condition) is shown in Supplementary Figure \ref{fig:s3}.

\subsection{Group interaction}

In the current version of the hardware/software platform, the total number of participants (HPs and/or VPs) needs to belong to the range $[3,7]$. Denoting with $N$ the total number of HPs and with $M$ the total number of VPs, $N+1$ machines are necessary to carry out these experiments.

\begin{figure}[h!]
\begin{center}
\includegraphics[width=11cm]{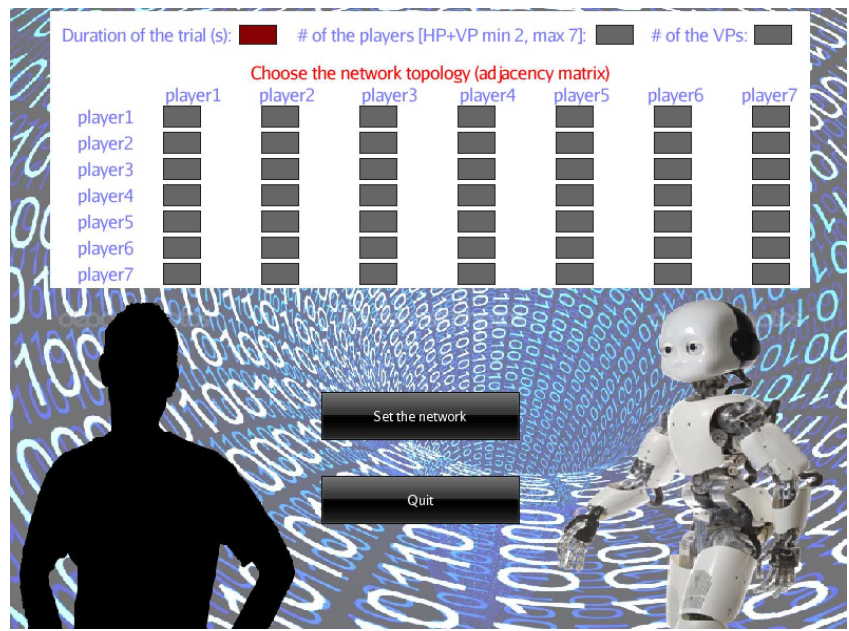}
\end{center}
\caption{\textbf{Screen for the administrator to select duration of the trial, number of players and set the network topology.} When setting the network, entering $1$ in position $(i,j)$ corresponds to allowing agent $i$ to see the motion of agent $j$; $0$ has to be entered in position $(i,i)$.}\label{fig:s4}
\end{figure}

\begin{figure}[h!]
\begin{center}
\includegraphics[width=13cm]{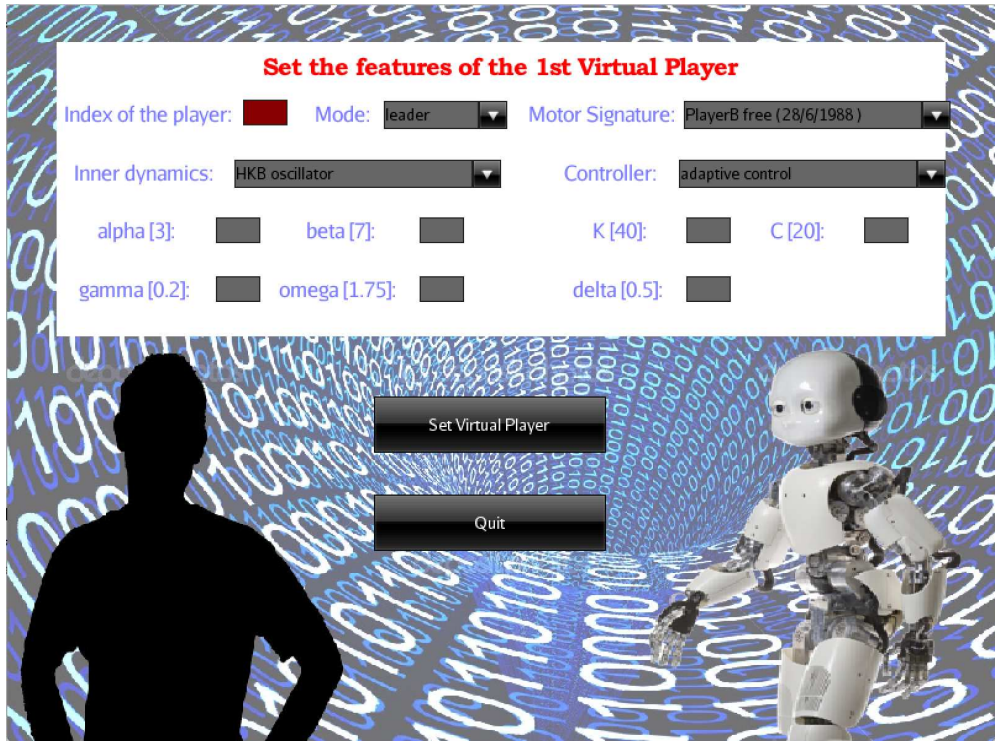}
\end{center}
\caption{\textbf{Screen for the administrator to set the VP in \emph{Group interaction} experiments.} For each VP, the administrator selects its mode (leader or follower), inner dynamics and control (if no parameters are specified, the default values in the brackets are used), motor signature (from the database) and an integer index uniquely identifying the virtual agent in the network.}\label{fig:s5}
\end{figure}

\begin{itemize}

\item For mixed HP-VP networks, the administrator:
\begin{enumerate}
\item runs the server module on a terminal window of a machine;
\item runs the database (MySql server) through \emph{MySql Workbench};
\item runs the administrator module on a different terminal window and selects \emph{Set network topology};
\item selects duration of the trial, number of HPs and VPs (between $3$ and $N+M$ combined) and sets the network topology (Supplementary Figure \ref{fig:s4});
\item for each VP, selects its mode (leader or follower), inner dynamics and control (if no parameters are specified, default values are used), motor signature (from the database) and an integer index (between $1$ and $N +M$) uniquely identifying it in the network (Supplementary Figure \ref{fig:s5}).
\end{enumerate}

\item For HP-HP networks, the administrator:
\begin{enumerate}
\item runs the server module on a terminal window of a machine;
\item runs the administrator module on a different terminal window and selects \emph{Set network topology};
\item selects duration of the trial, number of HPs (between $3$ and $N$) and sets the network topology (Supplementary Figure \ref{fig:s4}).
\end{enumerate}
\end{itemize}
When setting the network topology, entering $1$ in position $(i,j)$ corresponds to allowing agent $i$ to see the motion of agent $j$; $0$ has to be entered in position $(i,i)$.

\begin{figure}[h!]
\begin{center}
\includegraphics[width=13cm]{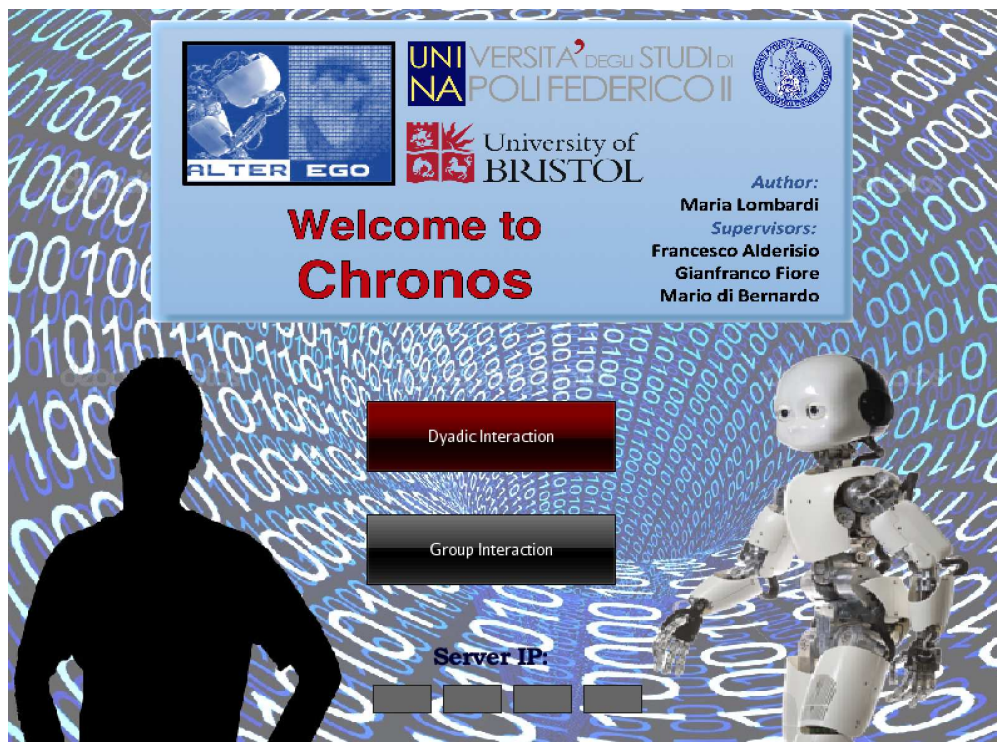}
\end{center}
\caption{\textbf{Screen for each human participant to choose the kind of trial to be performed.} The player needs to enter the IP address of the server in the appropriate text boxes.}\label{fig:s2}
\end{figure}

For both types of networks, all HPs:
\begin{enumerate}
\item connect their machine to the WLAN provided by the router;
\item run their client module on a terminal window of their own personal computer and, after entering the server's IP address, select \emph{Group interaction} (Supplementary Figure \ref{fig:s2}) and enter an integer index (between $1$ and $N+M$ in the case of HP-VP networks, between $1$ and $N$ in the case of HP-HP networks) uniquely identifying them in the network;
\item perform the experiment.
\end{enumerate}

An example of the gaming screen in the case of \emph{Group interaction} experiments among $4$ players is shown in Supplementary Figure \ref{fig:s6}.

\subsection{Data storage and file names}
For each player (HP or VP), all data can be saved on the server's machine by calling a shell script named \emph{saveData} from its terminal at the end of each trial (see link \url{https://dibernardogroup.github.io/Chronos} for more information). Data is saved as a \emph{.txt} file (motor signatures are stored in the database as well) made up of three (for \emph{Solo experiments}) or two (for all the other cases) columns: the first contains the time instants (in $ms$) in which data was sampled, the second contains the sampled position (in $dm$) and the third (only for \emph{Solo experiments}) the respective velocity (in $\frac{dm}{s}$).

In particular, each file is saved as \texttt{P}$N_p$\texttt{\_0}$N_t$\texttt{\_}$Z$\texttt{\_1d.txt}, where
\begin{itemize}
\item $N_p$ is the total number of players involved in the trial;
\item $N_t$ is an integer index identifying the trial;
\item $Z$ uniquely identifies the players involved in the trial.
\end{itemize}
In the case of \emph{Solo experiments}, a further parameter is added to identify the kind of motion (sinusoidal or free) performed by the human player.

For instance:
\begin{itemize}
\item[-] In \emph{Solo experiments}, \texttt{P1\_03\_Sample\_free\_1d.txt} refers to the third trial of a player called \texttt{Sample} who performed a free motion in isolation (\texttt{sinusoidal} instead of \texttt{free} if the motion performed was a sine wave);
\item[-] In \emph{Dyadic interaction (Leader-Follower condition)}, \texttt{P2\_03\_L\_1d.txt} refers to the third trial of the leader (\texttt{F} instead of \texttt{L} for the follower);
\item[-] In \emph{Dyadic interaction (Joint improvisation condition)}, \texttt{P2\_03\_JI1\_1d.txt} refers to the third trial of the player identified with index \texttt{1} (\texttt{JI2} instead of \texttt{JI1} for the other player);
\item[-] In \emph{Group interaction}, \texttt{P5\_02\_4\_1d.txt} refers to the second trial of the player identified with index \texttt{4} in a network of \texttt{5} participants.
\end{itemize}

\section{Supplementary Tables and Figures}
\label{sec:suppInf4}


\subsection{Tables}

\begin{center}
\captionof{table}{
{\bf Mean value $\mu \left( \rho_g \right)$ and standard deviation $\sigma \left( \rho_g \right)$ over time of the group synchronization index in the \emph{Group interaction} experiments, averaged over the $3$ trials for each topology -- Group 1.} This table shows $\mu \left( \rho_g \right) \pm \sigma \left( \rho_g \right)$ for the $5$ topologies of interest.\\} \label{tab:1}
    \begin{tabular}{| l | l |}
    \hline
    {\bf Topology} & {\bf All-to-all}\\ \hline
    $4$ HPs & $0.9689 \pm 0.0551$ \\ \hline
    $4$ HPs + $1$VP (leader) connected to $1$ HP & $0.8044 \pm 0.1346$\\ \hline
    $4$ HPs + $1$VP (leader) connected to $2$ HPs & $0.8057 \pm 0.1572$ \\ \hline
    $4$ HPs + $1$VP (leader) connected to $4$ HPs & $0.7915 \pm 0.1518$ \\ \hline
    $4$ HPs + $1$VP (follower) connected to $4$ HPs & $0.9331 \pm 0.0887$ \\ \hline
    \end{tabular}
\end{center}

\clearpage
\begin{center}
\captionof{table}{
{\bf Mean value of the dyadic synchronization indices $\rho_{d_{h,k}}$ over the $3$ trials for each of the $5$ topologies of interest in the \emph{Group interaction} experiments -- Group 1.} Player $5$ represents the additional VP.\\} \label{tab:2}
    \begin{tabular}{| l | p{1cm} |  p{1cm} | p{1cm} | p{1cm} | p{1cm} |}
    \hline
    {$4$ HPs -- Agents}& $1$ & $2$ & $3$ & $4$ & $5$ \\ \hline
    $1$ & - & 0.89 & 0.94 & 0.94 & - \\ \hline
    $2$ & 0.89 & - & 0.91 & 0.93 & - \\ \hline
    $3$ & 0.94 & 0.91 & - & 0.95 & - \\ \hline
    $4$ & 0.94 & 0.93 & 0.95 & - & - \\ \hline
    $5$ & - & - & - & - & - \\ \hline \hline
   {$4$ HPs + $1$VP (leader) connected to $1$ HP -- Agents}& $1$ & $2$ & $3$ & $4$ & $5$ \\ \hline
    $1$ & - & 0.94 & 0.90 & 0.91 & 0.12 \\ \hline
    $2$ & 0.94 & - & 0.90 & 0.91 & 0.15 \\ \hline
    $3$ & 0.90 & 0.90 & - & 0.87 & 0.08 \\ \hline
    $4$ & 0.91 & 0.91 & 0.87 & - & 0.17 \\ \hline
    $5$ & 0.12 & 0.15 & 0.08 & 0.17 & - \\ \hline \hline
   {$4$ HPs + $1$VP (leader) connected to $2$ HPs -- Agents}& $1$ & $2$ & $3$ & $4$ & $5$ \\ \hline
    $1$ & - & 0.85 & 0.88 & 0.85 & 0.22 \\ \hline
    $2$ & 0.85 & - & 0.86 & 0.93 & 0.22 \\ \hline
    $3$ & 0.88 & 0.86 & - & 0.84 & 0.24 \\ \hline
    $4$ & 0.85 & 0.93 & 0.84 & - & 0.22 \\ \hline
    $5$ & 0.22 & 0.22 & 0.24 & 0.22 & - \\ \hline \hline
   {$4$ HPs + $1$VP (leader) connected to $4$ HPs -- Agents}& $1$ & $2$ & $3$ & $4$ & $5$ \\ \hline
    $1$ & - & 0.87 & 0.88 & 0.82 & 0.16 \\ \hline
    $2$ & 0.87 & - & 0.83 & 0.83 & 0.20 \\ \hline
    $3$ & 0.88 & 0.83 & - & 0.80 & 0.15 \\ \hline
    $4$ & 0.82 & 0.83 & 0.80 & - & 0.29 \\ \hline
    $5$ & 0.16 & 0.20 & 0.15 & 0.29 & - \\ \hline \hline
    {$4$ HPs + $1$VP (follower) connected to $4$ HPs -- Agents}& $1$ & $2$ & $3$ & $4$ & $5$ \\ \hline
    $1$ & - & 0.82 & 0.78 & 0.82 & 0.88 \\ \hline
    $2$ & 0.82 & - & 0.82 & 0.85 & 0.90 \\ \hline
    $3$ & 0.78 & 0.82 & - & 0.86 & 0.91 \\ \hline
    $4$ & 0.82 & 0.85 & 0.86 & - & 0.94 \\ \hline
    $5$ & 0.88 & 0.90 & 0.91 & 0.94 & - \\ \hline
    \end{tabular}
\end{center}

\clearpage
\begin{center}
\captionof{table}{
{\bf Mean value $\mu \left( \rho_g \right)$ and standard deviation $\sigma \left( \rho_g \right)$ over time of the group synchronization index in the \emph{Group interaction} experiments, averaged over the total number of trials -- Group 2.} This table shows $\mu \left( \rho_g \right) \pm \sigma \left( \rho_g \right)$ for both undirected and directed topologies.\\} \label{tab:3}
    \begin{tabular}{| l | l | l |}
    \hline
    {\bf Topology} & {\bf Undirected} & {\bf Directed} \\ \hline
    Complete graph & $0.8888 \pm 0.1925$ & $0.8694 \pm 0.1616$ \\ \hline
    Ring graph & $0.8145 \pm 0.2203$ & $0.8322 \pm 0.1864$\\ \hline
    Path graph & $0.7814 \pm 0.2068$ & $0.8643 \pm 0.1611$ \\ \hline
    Star graph & $0.9028 \pm 0.1702$ & $0.9070 \pm 0.1451$ \\ \hline
    \end{tabular}
\end{center}

\begin{center}
\captionof{table}{
{\bf Mean value, over the total number of trials, of the dyadic synchronization indices $\rho_{d_{h,k}}$ in the undirected complete, ring, path and star graph in the \emph{Group interaction} experiments -- Group 2.} Indices related to players who were coupled to each other in the experiments (there exists an edge between them in the respective underlying topology) are represented in bold.\\} \label{tab:4}
    \begin{tabular}{| l | p{2cm} |  p{2cm} | p{2cm} | p{2cm} | p{2cm} |}
    \hline
    {Complete graph -- Agents}& $1$ & $2$ & $3$ & $4$ & $5$ \\ \hline
    $1$ & - & {\bf 0.86} & {\bf 0.90} & {\bf 0.73} & {\bf 0.73} \\ \hline
    $2$ & {\bf 0.86} & - & {\bf 0.89} & {\bf 0.72} & {\bf 0.77} \\ \hline
    $3$ & {\bf 0.90} & {\bf 0.89} & - & {\bf 0.73} & {\bf 0.77} \\ \hline
    $4$ & {\bf 0.73} & {\bf 0.72} & {\bf 0.73} & - & {\bf 0.81} \\ \hline
    $5$ & {\bf 0.73} & {\bf 0.77} & {\bf 0.77} & {\bf 0.81} & - \\ \hline \hline
   {Ring graph -- Agents}& $1$ & $2$ & $3$ & $4$ & $5$ \\ \hline
    $1$ & - & {\bf 0.84} & 0.75 & 0.58 & {\bf 0.58} \\ \hline
    $2$ & {\bf 0.84} & - & {\bf 0.85} & 0.57 & 0.58 \\ \hline
    $3$ & 0.75 & {\bf 0.85} & - & {\bf 0.60} & 0.61 \\ \hline
    $4$ & 0.58 & 0.57 & {\bf 0.60} & - & {\bf 0.74} \\ \hline
    $5$ & {\bf 0.58} & 0.58 & 0.61 & {\bf 0.74} & - \\ \hline \hline
   {Path graph -- Agents}& $1$ & $2$ & $3$ & $4$ & $5$ \\ \hline
    $1$ & - & {\bf 0.88} & 0.77 & 0.52 & 0.51 \\ \hline
    $2$ & {\bf 0.88} & - & {\bf 0.88} & 0.57 & 0.53 \\ \hline
    $3$ & 0.77 & {\bf 0.88} & - & {\bf 0.52} & 0.50 \\ \hline
    $4$ & 0.52 & 0.57 & {\bf 0.52} & - & {\bf 0.89} \\ \hline
    $5$ & 0.51 & 0.53 & 0.50 & {\bf 0.89} & - \\ \hline \hline
   {Star graph -- Agents}& $1$ & $2$ & $3$ & $4$ & $5$ \\ \hline
    $1$ & - & {\bf 0.89} & 0.94 & 0.70 & 0.82 \\ \hline
    $2$ & {\bf 0.89} & - & {\bf 0.92} & {\bf 0.69} & {\bf 0.88} \\ \hline
    $3$ & 0.94 & {\bf 0.92} & - & 0.69 & 0.82 \\ \hline
    $4$ & 0.70 & {\bf 0.69} & 0.69 & - & 0.75 \\ \hline
    $5$ & 0.82 & {\bf 0.88} & 0.82 & 0.75 & - \\ \hline
    \end{tabular}
\end{center}

\clearpage
\begin{center}
\captionof{table}{
{\bf Mean value, over the total number of trials, of the dyadic synchronization indices $\rho_{d_{h,k}}$ in the directed complete, ring, path and star graph in the \emph{Group interaction} experiments -- Group 2.} Indices related to players who were coupled to each other in the experiments (there exists an edge between them in the respective underlying topology) are represented in bold.\\} \label{tab:5}
    \begin{tabular}{| l | p{2cm} |  p{2cm} | p{2cm} | p{2cm} | p{2cm} |}
    \hline
    {Complete graph -- Agents}& $1$ & $2$ & $3$ & $4$ & $5$ \\ \hline
    $1$ & - & {\bf 0.67} & {\bf 0.69} & {\bf 0.61} & {\bf 0.76} \\ \hline
    $2$ & {\bf 0.67} & - & {\bf 0.84} & {\bf 0.70} & {\bf 0.80} \\ \hline
    $3$ & {\bf 0.69} & {\bf 0.84} & - & {\bf 0.81} & {\bf 0.78} \\ \hline
    $4$ & {\bf 0.61} & {\bf 0.70} & {\bf 0.81} & - & {\bf 0.72} \\ \hline
    $5$ & {\bf 0.76} & {\bf 0.80} & {\bf 0.78} & {\bf 0.72} & - \\ \hline \hline
   {Ring graph -- Agents}& $1$ & $2$ & $3$ & $4$ & $5$ \\ \hline
    $1$ & - & {\bf 0.80} & 0.66 & 0.64 & {\bf 0.74} \\ \hline
    $2$ & {\bf 0.80} & - & {\bf 0.79} & 0.55 & 0.60 \\ \hline
    $3$ & 0.66 & {\bf 0.79} & - & {\bf 0.57} & 0.49 \\ \hline
    $4$ & 0.64 & 0.55 & {\bf 0.57} & - & {\bf 0.81} \\ \hline
    $5$ & {\bf 0.74} & 0.60 & 0.49 & {\bf 0.81} & - \\ \hline \hline
   {Path graph -- Agents}& $1$ & $2$ & $3$ & $4$ & $5$ \\ \hline
    $1$ & - & {\bf 0.82} & 0.75 & 0.59 & 0.51 \\ \hline
    $2$ & {\bf 0.82} & - & {\bf 0.88} & 0.72 & 0.61 \\ \hline
    $3$ & 0.75 & {\bf 0.88} & - & {\bf 0.84} & 0.68 \\ \hline
    $4$ & 0.59 & 0.72 & {\bf 0.84} & - & {\bf 0.80} \\ \hline
    $5$ & 0.51 & 0.61 & 0.68 & {\bf 0.80} & - \\ \hline \hline
   {Star graph -- Agents}& $1$ & $2$ & $3$ & $4$ & $5$ \\ \hline
    $1$ & - & {\bf 0.86} & 0.90 & 0.70 & 0.77 \\ \hline
    $2$ & {\bf 0.86} & - & {\bf 0.94} & {\bf 0.76} & {\bf 0.84} \\ \hline
    $3$ & 0.90 & {\bf 0.94} & - & 0.74 & 0.83 \\ \hline
    $4$ & 0.70 & {\bf 0.76} & 0.74 & - & 0.80 \\ \hline
    $5$ & 0.77 & {\bf 0.84} & 0.83 & 0.80 & - \\ \hline
    \end{tabular}
\end{center}

\clearpage
\subsection{Figures}

\begin{figure}[h!]
\begin{center}
\includegraphics[width=17cm]{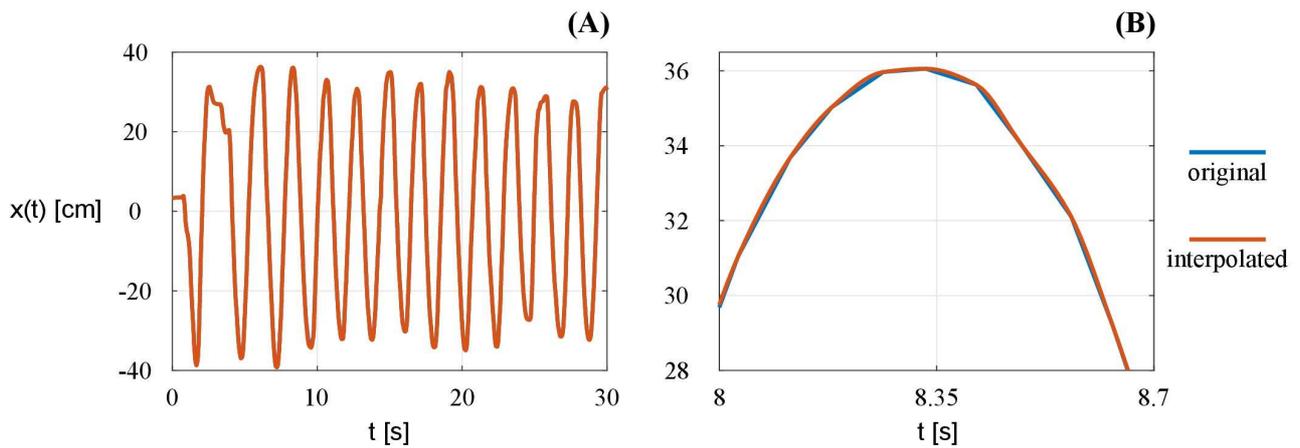}
\end{center}
\caption{\textbf{Cubic interpolation for a player's trajectory in \emph{Group interaction} -- Group 2.} (A) Original (blue line) and interpolated (orange line) position trajectory extracted from a trial of duration $30s$, and respective zoom (B), are shown. Data was originally stored with a frequency rate of $13Hz$, and then underwent cubic interpolation ($100Hz$).}\label{fig:s1}
\end{figure}

\begin{figure}[h!]
\begin{center}
\includegraphics[width=11cm]{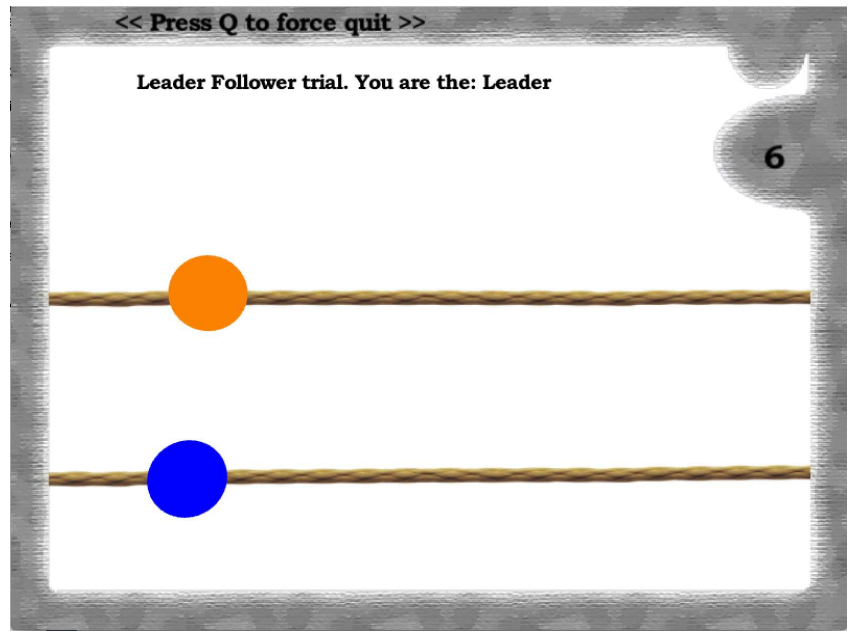}
\end{center}
\caption{\textbf{Gaming screen for human participants in \emph{Dyadic interaction} experiments (Leader-Follower condition).} Each participant sees her/his own motion represented by a blue circle, and that of the others s/he is interacting with represented by an orange circle. In this case, the gaming screen of the Leader is shown. Top-right: countdown timer (in $s$).}\label{fig:s3}
\end{figure}

\begin{figure}[h!]
\begin{center}
\includegraphics[width=11cm]{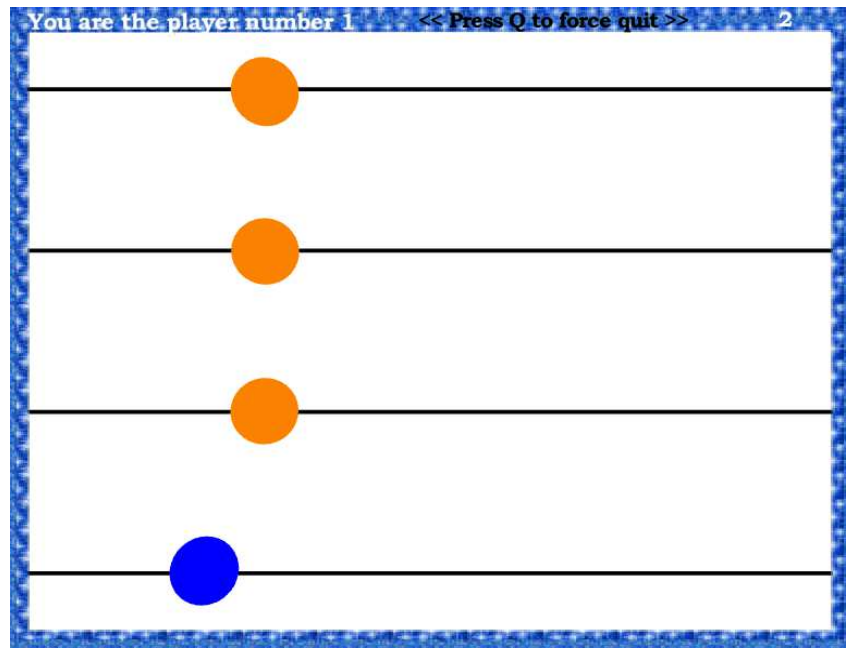}
\end{center}
\caption{\textbf{Gaming screen for human participants in \emph{Group interaction} experiments.} Each participant sees her/his own motion represented by a blue circle, and those of the others s/he is topologically connected with represented by orange circles, respectively. Top-left: index of the player in the network. Top-right: countdown timer (in $s$).}\label{fig:s6}
\end{figure}


\end{document}